\documentclass[fleqn,twoside]{article}
\usepackage{espcrc2}
\usepackage{graphicx}
\title{Thermal neutron flux produced by EAS at various altitudes}
\author{Yu.V. Stenkin,
\address{Insitute for Nuclear Research of Russian Academy of
Sciences \\} \thanks{email: (stenkin@sci.lebedev.ru)} 
V.V. Alekseenko \address{Insitute for Nuclear Research of Russian
Academy of Sciences\\},
D.M. Gromushkin \address{Moscow Physical Engineering Institute\\} %
Y. Liu \address{HeBei Normal University\\}, %
X.H. Ma \address{Insitute of High Energy Physics of
Chinese Academy of Sciences\\}\thanks{email: (maxh@ihep.ac.cn)} and 
J. Zhao \address{Insitute of High Energy Physics of Chinese
Academy of Sciences\\}}
\begin{document}

\begin{abstract}
The results of Monte-Carlo simulations of Extensive Air Shower are
presented to show the difference of hadronic component content at
various altitudes with the aim to choose an optimal altitude for the
PRISMA-like experiment. CORSIKA program for EAS
simulations with QGSJET and GHEISHA models was used to calculate the
number of hadrons reaching the observational level inside a ring of
50 m radius around the EAS axis.  Then the number of neutrons
produced by the hadronic component was calculated using an empirical
relationship between the two components. We have tested the results
with the ProtoPRISMA array at sea level, and recorded neutrons are close to the simulation results.
\end{abstract}
\maketitle

{\it Keywords: EAS, detector, neutrons, altitudes}\\

\section{Introduction}

A novel type of Extensive Air Shower (EAS) array (the PRISMA
project) has been proposed some years ago \cite{lab1} to study cosmic ray
spectrum and mass composition in the "knee" region. Existing
experimental data in the knee region contradict each other and new
approaches to the so-called "knee problem" are needed to solve this
complicated and old problem. The PRISMA project is based on an idea
that the main EAS component - hadrons has to be measured first of
all. Special detectors (en-detectors) have been developed for this
purpose. It was proposed to combine the central part of the PRISMA
array with the LHAASO project \cite{lab2} by introducing the en-detectors in
the center of LHAASO array. The en-detectors will make the LHAASO array
sensitive to the hadronic EAS component thus making it more powerful
and informative.

\section{Calculations}

We used the CORSIKA program \cite{lab3} (ver. 6.900) for EAS simulations
with QGSJET and GHEISHA models. Calculations were performed
for two primaries: proton and iron and for
two altitudes: near sea level (170 m a.s.l.) and high mountain 4300
m a.s.l.  As a first step we made calculations for fixed primary
energies from 10TeV through 10PeV and zenith angles $0-45^{0}$. We
present here results for the number of hadrons and produced by them
thermal neutrons inside a ring of 50 m radius around the EAS axis as
a function of primary energy. The distributions over these numbers
are also obtained.

As any other Monte-Carlo program for EAS simulation, CORSIKA can not
process particles with very low energies. We used following cuts for
particles: 50MeV for hadrons, 0.5GeV for muons, 60keV for electrons
and gammas. The mean number of evaporation neutrons $<n>$ produced by
hadrons in 3-m layer of surrounding soil and/or construction
materials has been calculated using an empirical relationship
between them and the parent hadron energy (in GeV):
\begin{equation}
<n>{\approx}36{\cdot}E^{0.56}_{h}
\end{equation}
This relationship originated from secondary particles production in
hadronic interactions and was obtained taking into account
experimental data \cite{lab4} and atomic mass A dependence of neutron
production: $<n> {\sim} A^{0.4} $\cite{lab5}. Therefore, the total number of
produced secondary neutrons should be summarized over all hadrons:
\begin{equation}
<n_{tot}>={\sum}36{\cdot}E^{0.56}_{h}
\end{equation}
It is easy to see that due to slow dependence on hadron energy and
due to slow change of hadron mean energy with primary energy, $eq.
2$ can be simplified to:
\begin{equation}
<n_{tot}>{\approx}{\sum}36{\cdot}<E_{h}>^{0.56}{\approx}36{\cdot}N_{h}{\cdot}<E_{h}>^{0.56}
\end{equation}
The latter means that the total number of evaporation neutrons produced in EAS should be more or less proportional to the
number of high-energy hadrons reaching the observation level. The
great bulk of these neutrons are thermalized later. Thus recording
thermal neutrons by detectors spread over big enough area one could
recover the number of hadrons in EAS.  This idea is a basis of the
PRISMA project.

\section{Simulation Results}
Mean numbers of produced neutrons and hadrons inside a ring of 50 m
as a function of primary energy are shown in Fig. 1 for primary
proton and iron and for near sea level. As one can see, all
dependencies can be fitted with power law functions. All indices for protons
are close to 1: ${\sim}1.14$ at sea level and ${\sim}1.05$ at 4300 m. For iron the indeces are little bit steeper at these energies. Small difference between primary proton and iron at the highest energy
makes us sure that recovery of primary energy would be more adequate in the knee region.
One can also see that the mean number of neutrons at sea level and at 4300 m a.s.l. differ by a factor of $\sim 10$.
The latter means that at Yangbajing level the array threshold energy can be lower by a factor of $10^{1/1.7}\approx 4$.
\begin{figure}
\begin{center}
\includegraphics[width=8.5cm]{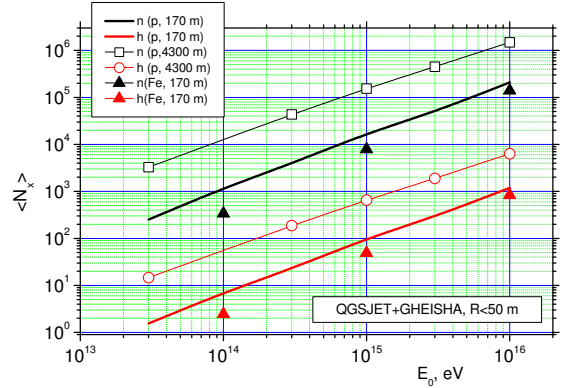}
\caption{\label{fig1}  Mean numbers of hadrons and secondary
evaporation neutrons inside ring of 50 m radius as a function of
primary energy at sea level for proton and iron primaries and at
4300 m for primary proton.}
\end{center}
\end{figure}
\begin{figure*}[htp]
\begin{center}
\includegraphics[width=16cm]{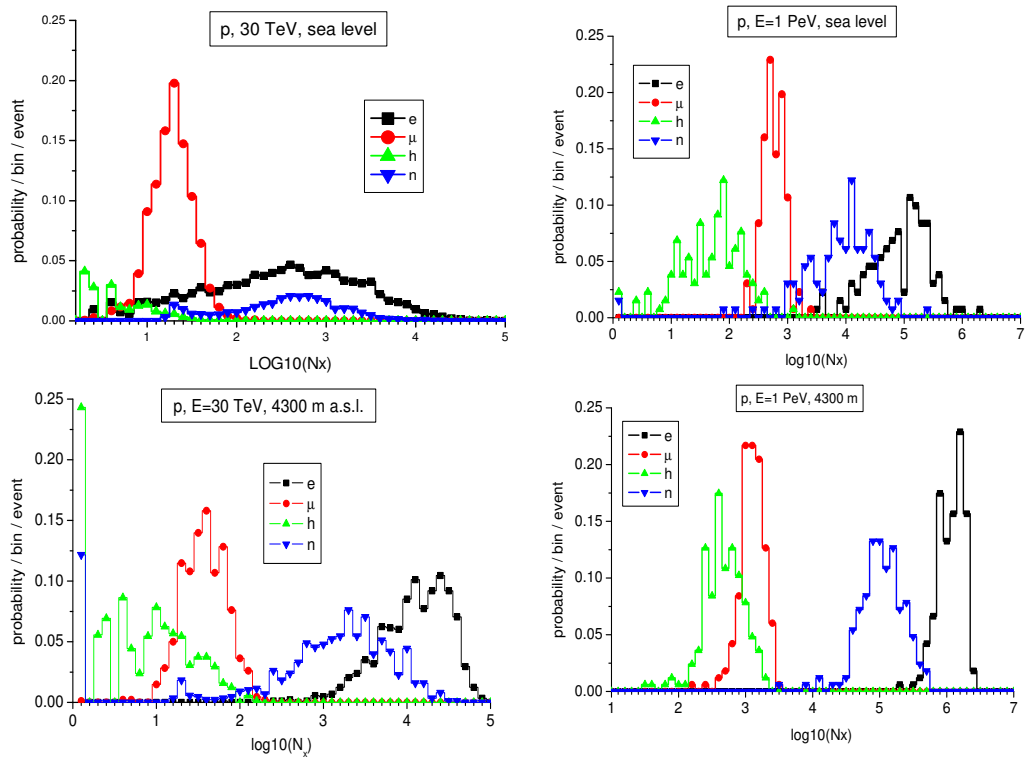}
\caption{\label{fig2} Different EAS components size distributions
for primary protons and primary energies equal to 30 TeV (left
panel) and 1 PeV (right panel) for sea level (upper panel) and for
4300 m a.s.l.(bottom panel)}
\end{center}
\end{figure*}
There are shown in Fig. 2 the distributions of different EAS
components numbers inside 50 m around the axis for 2 primary
energies: 30TeV and 1PeV. An interesting issue here is rather high mean number of produced neutrons in comparison with hadrons and muons. The latter means that thermal neutrons could be a better primary energy estimator than muons or high energy hadrons. Figure 2 also presents that at high altitude, within the primary energy range 30TeV-10PeV, the secondaries have higher quantities and then lead to higher energy resolution than ones at sea level. It is necessary to test these simulation results at different altitudes.

\section{Test at sea level}

\subsection{The ProtoPRISMA array}

We have made several tests with the PRISMA prototypes in Moscow.
Our test results are consistent with the simulations. A prototype
of the PRISMA project array (ProtoPRISMA)
\cite{lab6} has been
developed and started running on a base of the NEVOD-DECOR
detector at National Research Nuclear University MEPhI in Moscow
(170 m a.s.l.). It consists now of 32 inorganic scintillator
$\it{en-detectors}$ situated inside the experimental building at a
level of 4th floor just around the NEVOD water pool. The neutron
recording efficiency of the en-detector is equal to $\epsilon_{d}
\approx 20\%$. The detectors have a cylindrical shape with the
scintillator area equal to 0.36 $m^{2}$. It is now our standard
en-detector made on a base of commercial polyethylene (PE) water
tank of 200 liters volume. The scintillator thin sheets are
situated on its bottom and are viewed by a single 6'' PMT
(FEU-200). The scintillator compound ZnS(Ag)+ LiF enriched with
$^{6}Li$ up to $90\%$ is very effective scintillator for heavy
particle detection. It produces 160000 photons per a neutron
capture through the reaction
 $^{6}Li$(n,a)t + 4.8 MeV. It allows us to collect more than 50 photoelectrons
from PMT photo-cathode per n-capture. Due to thin scintillator
layer (30 $mg/cm^{2}$), a single relativistic charged particle
produces very small signal. But, in a case of EAS passage,
correlated signals from many particles are summarized and can be
measured. Therefore, the same detectors are used to measure two
EAS components: hadronic (neutrons) and electromagnetic. That is
why we call them as en-detectors (electron-neutron). The array
consists now of 2 clusters of 16 detectors each. The clusters have
their own triggering system and work independently. Coincidence of
2 or more hit detectors within time of 2 $\mu$s produces a simple
trigger and on-line program preprocesses data, selects EAS events
and marks them in accordance with different mathematical trigger
conditions. All pulses are integrated with a time constant of 1
$\mu$s and are digitized using 10 bits flash ADC with a step of 1
$\mu$s. In a case of powerful enough EAS, full pulse waveform of
20 ms duration is stored. For energy deposit measurement in a wide
range we use 2 signals from each detectors: from the last 12th
dynode and from an intermediate 7th dynode as well. Due to this we
have the energy range more than $10^{4}$. Delayed pulses from
thermal neutrons captures are measured in each detector in the
time gate 0.1 - 20 ms.

In order to test our simulations at sea level, we have processed the experimental data accumulated for 2 winter months with the aim to check eq.(1) normalization factor. It is obvious that absolute value of neutron yield depends on the experimental conditions. In our case the array is not situated in open air, but inside the building. Therefore, the result can be regarded as an estimation only. Detailed neutron yield measurements will be performed later when the next  PRISMA prototype will be running in Yangbajing area in real conditions.

\subsection{Experimental results}

Mean number of recorded neutrons as a function of EAS size ($N_{e}$) is shown in fig. 3
in comparison with full-scale Monte-Carlo simulations made using CORSIKA (with the parameters mentioned above) and the ProtoPRISMA array simulations.
 For neutron production we use a formula similar to eq.(1)
 but with a normalization factor taking the our neutron recording efficiency into account.
 The efficiency can be estimated as follows:
 $\epsilon \approx \epsilon_{d}\cdot\epsilon_{em}\cdot\epsilon_{area}$, where $\epsilon_{em} \approx 0.04$
 is a probability for neutrons produced in soil (concrete) to escape to air
 \cite{lab7}
 and $\epsilon_{area}$=0.028 is a ratio of total detectors area (13.5 $m^{2}$) to the array area
 (484.5 $m^{2}$)(coverage).
 Taking these efficiencies into account one can transform eq.(2) to:
\begin{equation}
<n_{tot}>{\approx C\cdot\epsilon\cdot}36{\cdot}E^{0.56}_{h}{\approx
C\cdot}0.0080{\cdot}E^{0.56}_{h}
\end{equation}
This relationship was used in the simulations of the experiment.
The normalization factor (C) was varied to adjust it to the
experimental points. Points in fig.3 were obtained with C=0.625.
Agreement is rather well (note that due to limited statistics for
the highest EAS energies corresponding calculated points at the right tail are
little bit lower).
\begin{figure}
\begin{center}
\includegraphics[width=8.5cm]{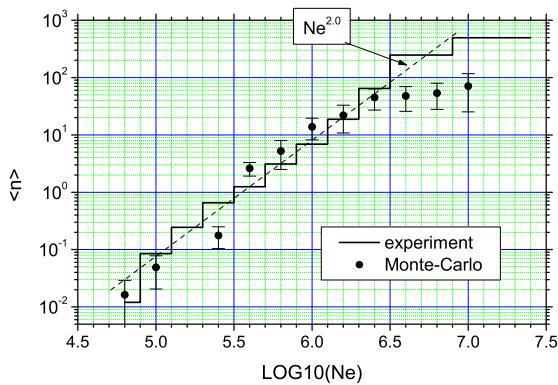}
\end{center}
\caption{\label{fig3} Mean number of recorded neutrons as a
function of EAS size for Moscow site. Histogram - experiment,
points - simulations.}
\end{figure}
As one can see, the mean number of recorded neutrons in the
ProtoPRISMA experiment is $\sim$ 1.6 times lower than expected
one. The reason could be found in the experimental details:
detectors are placed inside building (not in open air), the
structure of the array is asymmetrical and the detector coverage
in 2 clusters are different
\cite{lab6} and finally, an
existence of a large water pool inside the array affects
undoubtedly on the neutrons yield making it lower.

\section{Summary}

In the frames of Russian-Chinese collaboration Monte-Carlo EAS
simulations have been performed and absolute normalization to the
experiment has been done. We obtained rather low yield of neutrons
at the ProtoPRISMA location in Moscow due to specifics of the
experiment. Nevertheless, even at such yield and even at sea level
it is possible to obtain some new results using the novel type of
EAS array. We are looking forward to make such experiment at high
altitude in Yangbajing region. It would make us possible to check
the method at high altitude, to make calibration using the
ARGO-YBJ facilities and to measure real neutron yield in the
future experiment site. The present calculations show that the
number of produced neutrons at Yangbajing level is by a factor of
10 higher than that at sea level in EAS of equal energy.

\section{ Acknowledgments}

Authors are grateful to Russian Foundation for Basic Research
(grants 09-02-12380-ofi-m and 11-02-01479a), to the ''Neutrino
program'' of Russian Academy of Sciences, the Federal Target
Program ''Scientific and educational cadres for innovative Russia''.
Authors are grateful to the National Natural Science Foundation of China (NSFC) ( the grant No. 10835007.)

Special thanks to the Computational Cluster of the Theoretical
division of INR RAS where numerical calculations were performed
and namely to Grigori Rubtsov for his help and discussions.

\end{document}